\documentclass[reprint,amsmath,amssymb,aps,pre,superscriptaddress]{revtex4-1}
\usepackage{graphicx}
\usepackage{grffile}
\usepackage{dcolumn}
\usepackage{bm}
\usepackage{hyperref}
\usepackage{color}

\begin{document}

\preprint{APS/123-QED}

\title{Stability of chiral crystal phase and breakdown of cholesteric phase in mixtures of active-passive chiral rods
}

\author{Jayeeta Chattopadhyay} 
\thanks{These authors contributed equally to this work}
\affiliation{%
Centre for Condensed Matter Theory, Department of Physics, Indian Institute of Science, Bangalore 560012, India
}%
\affiliation{Niels Bohr Institute, University of Copenhagen, Blegdamsvej 17, Copenhagen 2100, Denmark}
\author{Jaydeep Mandal}
\thanks{These authors contributed equally to this work}
\affiliation{%
Centre for Condensed Matter Theory, Department of Physics, Indian Institute of Science, Bangalore 560012, India
}%
\author{Prabal K. Maiti  }
\email{Corresponding author: maiti@iisc.ac.in}
\affiliation{%
Centre for Condensed Matter Theory, Department of Physics, Indian Institute of Science, Bangalore 560012, India
}%

\date{\today}

\begin{abstract}
In this study, we aim to explore the effect of chirality on the phase behavior of active helical particles driven by two-temperature scalar activity. We first calculate the equation of state of soft helical particles of various intrinsic chiralities using molecular dynamics (MD) simulation. In equilibrium,  the emergence of various liquid crystal (LC) phases such as nematic ($N$), cholesteric ($N_{c}^{*}$), smectic ($Sm$) as well as crystal ($K$) crucially depends on the presence of the walls which induce homeotropic alignment. Next, we introduce activity through the two-temperature model: keep increasing the temperature of half of the helical particles (labeled as `hot' particles) while maintaining the temperature of the other half at a lower value (labeled as `cold' particles). Starting from a homogeneous isotropic ($I$) phase, we find the emergence of 2-TIPS: two temperature-induced phase separations between the hot and cold particles. We also observe the cold particles undergo an ordering transition to various LC phases even in the absence of a wall. This observation reveals that the hot-cold interface in the active system plays the role of a wall in the equilibrium system by inducing an alignment direction for the cold particles. However, in the case of a cholesteric phase, we observe activity destabilizes the $N_{c}^{*}$ phase by inducing smectic ordering in the cold zone while isotropic structure in the hot zone. The smectic ordering in the cold zone eventually transforms to a chiral crystal phase at high enough activity.

\end{abstract}
\maketitle

\section{\label{sec:level1} Introduction}
Understanding how chirality spreads from the molecular scale to the macroscopic scale is important due to its immense contribution to basic science and industrial applications. Chiral particles that only interact sterically are one particularly intriguing example. Helix is one of the most basic models in this class that gives rise to the cholesteric phase (or Chiral nematic)  \cite{de-gennes, marjolein-natcomm-2016,hima-JCP-2017}, in which 
particles undergo a rotation around a specific helical axis. The wavelength associated with this rotation is called the pitch of the macroscopic cholesteric phase $P_{chol}$. This wavelength can have values that are many orders of magnitude larger than the molecular size. Cholesteric phases can be observed in different systems like DNA \cite{nakata2007end,zanchetta2010right,fraccia2016liquid}, fd viruses(\cite{dogic2006ordered}) amyloid fibrils (\cite{nystrom2018confinement}) and others \cite{werbowyj1976liquid,gilbert1983liquid,tombolato2006chiral}. It has applications on wide areas of optoelectronic technology \cite{collings2002liquid}.
Another important aspect of the cholesteric phase is the sense of rotation or handedness. It is important to study the relation between the intrinsic helicity of the molecules and the macroscopic chiral phase. Right-handed DNA forms left-handed chiral macrophase \cite{livolant1988circular}. Left-handed filamentous virus are observed to form right-handed cholesteric phase \cite{tombolato2006chiral}. 
Liquid crystal ordering can be seen in ultrashort double-stranded DNA \cite{PhysRevE.95.032702} and RNA \cite{naskar2020liquid}, which are right-handed helices.

Different theoretical and computational models have been proposed to understand the link between the molecular and macroscopic chirality. Theoretical studies have been used to predict the pitch of the cholesteric phase \cite{straley1976theory}. Odijk \cite{odijk1987pitch} has studied the pitch of the cholesteric phase using particle models where rod-like particles are enveloped by thin chiral threads. Using the corkscrew model, Pelcovits \cite{pelcovits1996cholesteric} showed that the pitch of the cholesteric phase is independent of the flexibility of the molecule, but they do depend on the intrinsic chirality and the concentration. Similar recent studies involving hard helices \cite{fejer2011self} \cite{kolli2014communication,hima-softmatter-2014, marjolein-PRE-2014} explore the different phases obtained for various different molecular chiralities. Dussi et. $al.$ \cite{dussi2016entropy} have used twisted polyhedral-shaped shaped hard-particles and obtained entropically driven stable chiral nematic phases along with other novel phases. Cholesteric phases are difficult to obtain in numerical simulations as advanced methodologies and large simulation boxes are needed in order to correctly replicate the macroscopic pitch. Hence, many attempts have been made on that aspect as well. Cinacchi et al. \cite{hima-JCP-2017} have studied the cholesteric phase by taking the hard helices model using both Monte Carlo (MC) and Molecular Dynamics (MD) simulations and observed the existence of the cholesteric phase in a confined system. Wu et al. \cite{wu-softmatter-2018,wu-PRE-2019} have used a flexible chain model where chiral centres are helically arranged on the surface of backbone beads, and showed that the pitch of the resultant cholesteric phase depends on the chirality and flexibility of the molecules. Similar results have been observed by Tortora et al. \cite{tortora2018incorporating}. Therefore, the study of the interplay between the microscopic and macroscopic helicity and the underlying interactions is a fascinating field of research.\\

Interesting traits in spatial ordering and persistent dynamics for the systems of chiral particles appear when activity is incorporated in the system \cite{sevilla2016diffusion,huang2020dynamical}. The presence of wall on the system of chiral active particles is shown to have a rich array of diverse effects \cite{caprini2019active}. These phenomena appear due to one characteristic of the system: activity.  
Active matter is driven out of equilibrium by a constant supply of free energy, which it consumes either from the ambient environment or from its own mechanism and dissipates it by performing mechanical work \cite{Sriram1,Sriram2,redner-baskaran}. In most of the cases, activity is \textit{vectorial} in nature due to the force of self-propulsion. However, many physical and biological processes (like chromatin separation in nucleus \cite{ganai}, phase separation in colloidal systems) are governed by unequal sharing of available energy, thus \textit{scalar} in nature. Such systems can be modeled simply by assigning different diffusivity between the constituents of the system \cite{frey} or by coupling it to two different thermostats \cite{joanny-2015,joanny-2018,joanny-2020}. Such `two-temperature model' often results in a phase separation phenomena termed as 2-TIPS \cite{tips_mips} and has been used extensively for the last few years in a diverse setting of systems from Lennard-jones (LJ) particles \cite{tips_mips}\cite{Siva}, polymer \cite{kremer1,kremer2}, dumbbells \cite{nayana-arxiv-2021} to our previous works in rod-like particles \cite{Active-SRS, Ons-JC}. One of the aims of this study is to explore the effect of the two-temperature model in a system of soft chiral particles. \\

To validate and benchmark our model of soft helices, we first investigated the equilibrium phase behavior of a system of helical particles with different chirality. This allows us to study the dependence of macroscopic pitches on the density and intrinsic pitch of the constituent chiral particles in the presence and absence of a wall in the system. Having established the correct equilibrium phase behavior of the system, we also examined the behavior of a mixture of active and passive helical particles driven by two-temperature scalar activity. The mixture of particles shows phase separation. The particles in the cold domain experience an ordering transition from isotropic to nematic to the smectic and crystalline structure, whereas the particles in the hot domain remain in the isotropic phase, even when we start from an ordered structure. However, starting from a cholesteric phase, we observe activity destabilizes the $N_{c}^{*}$ phase by inducing smectic ordering in the cold zone while forming an isotropic structure in the hot zone. At high enough activity, the smectic ordering in the cold zone eventually transforms into a chiral crystal phase.\\

The paper is organized as follows: In section \ref{sec:level1}, we introduce the model and details of the simulation methods and techniques used. In section \ref{sec:level2}, we report various results from equilibrium simulations. We then report the results with the non-equilibrium results-- namely the behaviour of a system of active and passive chiral particles, which show the resulting phase separation phenomena (2-TIPS). There is also an ordering transition occurring in the cold zone. Finally, conclusion is drawn with the main results and future outlook of our work is shown in the section \ref{sec:level3}

\section{\label{sec:level1}Model and Simulation method}

\begin{figure}[]
    \centering
    \includegraphics[height=0.6\linewidth]{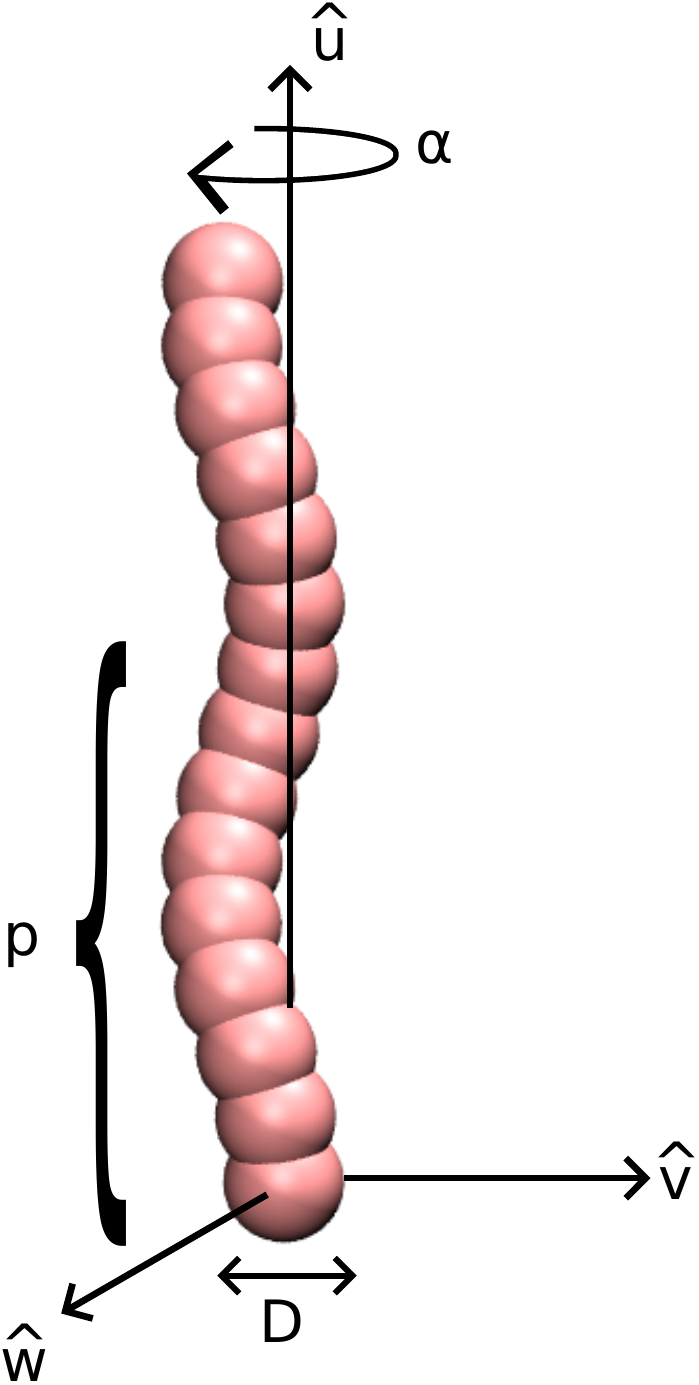}
    \caption{ Figure shows the schematic details of the left-handed soft helical particles constructed using the beads. The position of the beads rotate around a major axis ($\hat{u}$), in a clockwise fashion as shown by the angle $\alpha$ in the figure, which renders the intrinsic chirality of the molecule to be left handed. The $\hat{v}$ and $\hat{w}$ are the minor or short axis. The pitch $p$ of the particle is the wavelength associated with the helical arrangement of the position of the beads, as shown in the figure. The diameter $D$ of the beads is taken to be $1.0$ for our calculations. }
    \label{fig:schematic}
\end{figure}

\begin{figure}[]
    \centering
    \includegraphics[width=1\linewidth]{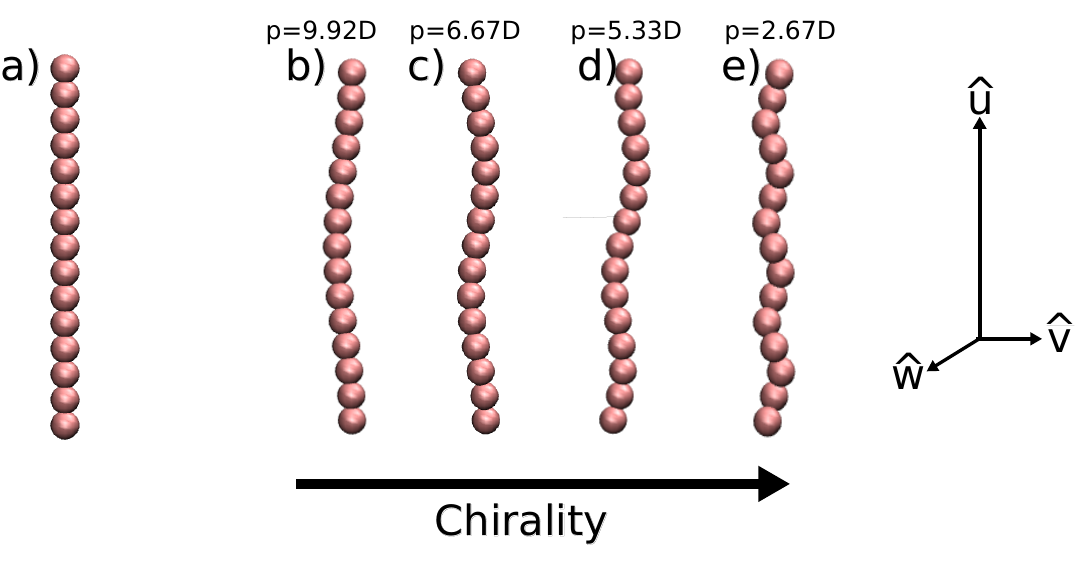}
    \caption{Schematic diagram of a straight rod made of soft beads in a). The structures of the left-handed helices of different intrinsic chiralities are shown in b),c),d), and e) for the molecular pitches $p = 9.92D,6.67D,5.33D,2.67D$ respectively, where $D$ is the diameter of one bead. Its long axis is defined by $\hat{u}$ and short axes by $\hat{v}$, $\hat{w}$ respectively. }
    \label{fig:mol-details}
\end{figure}

Helices can be modeled as a collection of a number of partially overlapping beads of diameter $D$ arranged rigidly around a helical axis $\hat{u}$, which is denoted as the long axis of the molecule (Fig. \ref{fig:schematic}). Our system is made of $N$ such left-handed helical particles of fixed contour length $L = 10D$. The chirality of a molecule can be defined by the pitch $p$ (the distance after which one bead makes a full rotation around the helical axis) and radius, $r$ varying which we can generate different helical particles ranging from straight rods to coils (Fig. \ref{fig:mol-details}). Our model is similar to that of Kolli \textit{et al.} \cite{hima-JCP-2013,hima-softmatter-2014} A schematic diagram of the model has been presented in Fig. \ref{fig:mol-details}. 

In our system, two beads belonging to two different particles interact via the repulsive part $U(r_{ij})$ of the Lennard-Jones (LJ) potential defined as \cite{weeks1971}:
\begin{align}
    U(r_{ij}) = \begin{cases}
            4\varepsilon\left[ \left( \frac{D}{r_{ij}} \right)^{12} - \left( \frac{D}{r_{ij}} \right)^6 \right] +\varepsilon, &r_{ij} < 2^{1/6} D \\
            0 ,&r_{ij} \geq 2^{1/6} D ,
        \end{cases}
\end{align}
where $r_{ij}$ is the distance between the $i^{th}$ and $j^{th}$ atom belonging to two different particles, and $\varepsilon$ is the unit of energy and denotes the strength of the interaction.

For convenience, the thermodynamic and structural quantities are scaled by the system parameters $\varepsilon,\ D$ and calculated in reduced units: temperature $T^* = k_BT/\varepsilon$, pressure $P^* = P/(k_BT)$, packing fraction $\eta = \rho v_0$, where $\rho = N/V$ is the number density and $v_0$ is the effective volume of the helical particle \cite{hima-JCP-2013}.

Obtaining cholesteric phases in computer  simulations
using the periodic boundary condition (PBC) is a difficult job, as the length scale of the $N_c^*$ phase may be incommensurate with the dimension of the box. To overcome this issue, we adopt a similar procedure followed by Kolli \textit{et al.} \cite{hima-JCP-2017} by inserting a wall along a specific direction of the simulation box. The presence of wall makes the system non-periodic along the wall direction.  We have inserted a wall on both sides of the $\hat{x}$ direction, which helps the system to align and exhibit  different liquid crystalline phases. The interaction between the wall and any bead of the helical particles is taken to be of the following form 
\begin{align}
    U(r) = \begin{cases}
            \varepsilon\left[\frac{2}{15} \left( \frac{D}{r} \right)^{9} - \left( \frac{D}{r} \right)^3 \right] +\varepsilon, &r < 0.4^{1/6} D \\
            0 ,&r
            \geq 0.4^{1/6} D ,
        \end{cases}
\end{align}
where $r$  is the distance between the wall and any bead. 


We built a system of $N=2000, 10000$ particles in a cubic simulation box using the PACKMOL software \cite{packmol1,packmol2}. Our simulation is done in the following way: First, we equilibrate the system at a certain density. Then insert wall along one direction (x) taking care to remove those particles whose constituent beads overlap with the wall. This lowers the density of the initial built system. We equilibrate  such system for 0.2 million (M) MD steps in NVT ensemble. 
 Finally, we increase the pressure of the confined system slowly under NPT ensemble using a semi-isotropic barostat i.e allowing the simulation box to fluctuate only in the y-z plane. The MD simulations in the NPT ensemble are done using the LAMMPS software \cite{lammps}. We use rigid body dynamics to integrate the equation of motion with a time step of $\Delta t = 0.001$ in reduced units (in units of $D\sqrt{m/\epsilon}$).
The temperature of the system is controlled by the Noose-Hoover thermostat \cite{nose1984molecular,hoover} with the temperature relaxation time $\tau_{T} = 100*\Delta t$ and pressure is controlled by the semi-isotropic Noose-Hoover barostat with the pressure relaxation time $\tau_{P} = 1000*\Delta t$. We run up to 50M MD steps to equilibrate the system and another 10 million to calculate the thermodynamic quantities.\\

 
The cholesteric phase is quantified by
calculating the macroscopic pitch $P_{chol}$. We divide the simulation box along the wall-direction x in a number of slabs and calculate the local nematic director $\hat{n}(x)$ for each of the slabs, which is the eigenvector corresponding to the largest eigenvalue of the traceless symmetric tensor $Q$ defined as:
\begin{equation} \label{eqn:nematic-order-parameter}
    Q_{\alpha \beta} = \frac{1}{N}\sum\limits_{i=1}^{N} \frac{3}{2} u_{i \alpha} u_{i \beta} \ -\ \frac{1}{2}\delta_{\alpha \beta}.
\end{equation}
In the above expression, $u_{i\alpha}$ is the $\alpha^{th}$ component of the long axis of the $i^{th}$ particles. The pitch was calculated by fitting the value of $|\hat{n(0)}\cdot\hat{n(x)}|$ with the function $|Cos(qx)|$ where the cholesteric pitch is defined as $P_{chol} = 2\pi/q$. 

Next, we incorporate activity in the system, using the two-temperature model. We introduce activity by choosing half of the helices randomly and assigning a higher temperature to them, while keeping the other particles' temperature fixed at a lower value equal to that of the initial equilibrium system. Let $ T_{h}^{*} $ and $ T_{c}^{*} $ be the temperature of the hot and cold particles, respectively. Initially, we equilibrate the system at  $T_{h}^{*} = T_{c}^{*} = 1.00 $, then increase $T_h^*$ in small steps upto $T_h^* = 20.0$, allowing the system to reach a steady state after each increase in $ T^*_h $, keeping the \textit{volume} of the simulation box constant throughout the simulation.  


We parameterize activity by 

\begin{equation}
\chi = \dfrac{ T_{h}^{*} -  T_{c}^{*}}{ T_{c}^{*}}
\end{equation}

For the active case, i.e., for $\chi \neq 0$, we choose the thermostat relaxation time $\tau_{T} = 0.01$ 
with an integration time-step $\Delta t = 0.001$ for both types of particles. We run the simulation for 3M to 4M integration time steps to reach steady state and another 1M steps to calculate thermodynamic and structural quantities.

\section{\label{sec:level2} Results and Discussion}

\begin{figure} [!htb]
	\centering
	\includegraphics[width= 0.7\linewidth]{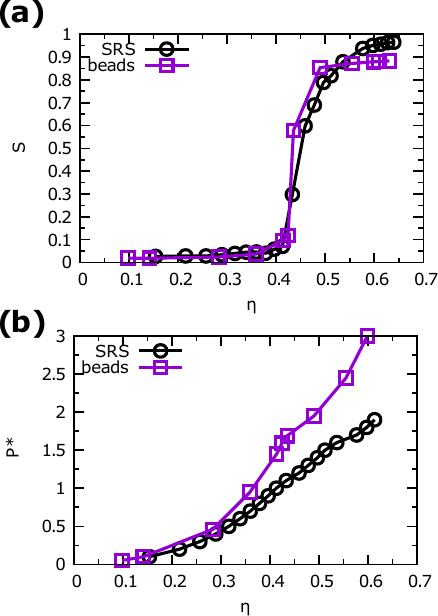}
	\caption{
    The phase behavior of rods made of soft beads with an effective aspect ratio $A_{eff} = 4.8$ is compared to that of soft repulsive spherocylinders. (a) Nematic order parameter $S$ and (b) reduced pressure $P^* = P/k_BT$ is plotted as a function of packing fraction $\eta$ for these two models that match quite well.
 }
    \label{fig:nem-eta-straight-rod}
    \end{figure}
\subsection{Equilibrium properties of soft helical particles}
\subsubsection{wall vs no-wall}

\begin{figure} [!htb]
	\centering
	\includegraphics[width= 0.7\linewidth]{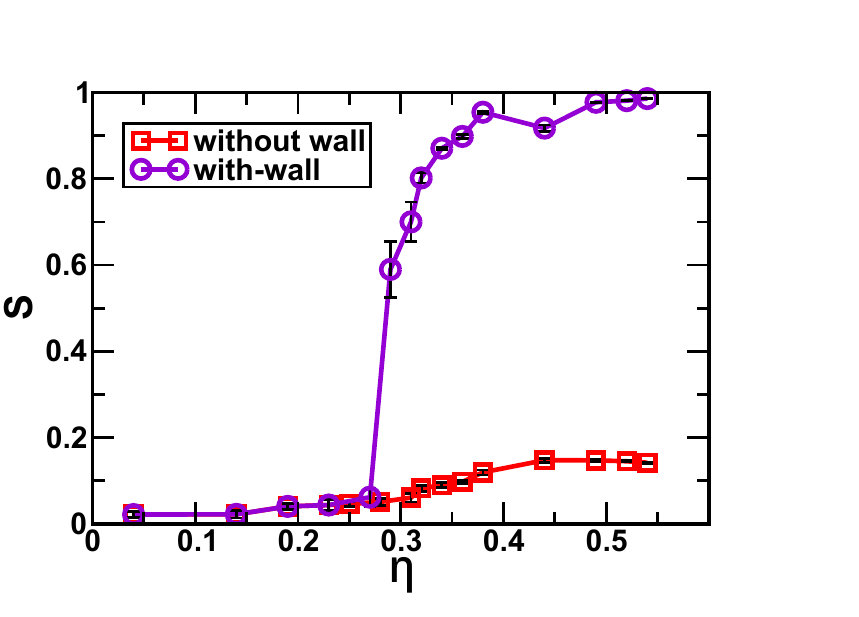}
	\caption{Nematic order parameter $S$ vs packing fraction $\eta$ for the pitch $p = 6.67D$ with and without wall. We observe that, in the absence of the wall, the system gets stuck into a jammed state; hence, the magnitude of the nematic order parameter is very low. Presence of a wall induces a homeotropic alignment that gives rise to ordered liquid crystal phases. Hence, the magnitude of $S$ increases with $\eta$.} \label{fig:nem-wall-com-p6.67}
	
\end{figure}

We simulate the system of soft helical particles under periodic boundary condition (PBC) using Molecular Dynamics (MD) simulations and find that, in the absence of the wall, the system gets stuck into a jammed state \ref{fig:without_wall_jammed_snapshots}, which makes it challenging to get liquid crystal ordering within our simulation time. This problem is overcome by inserting  a wall that induces the homeotropic alignment, which gives rise to the ordered LC phases. 
In Fig. \ref{fig:nem-wall-com-p6.67}, we show the effect of the wall in the phase behavior of the system with molecular pitch $p = 6.67D$.
It is observed that, the nematic order parameter is very low in the absence of a wall, while it increases with packing fraction in the presence of a wall. Hence, the equilibrium phase behavior for different molecular pitches are studied in the presence of the wall only.
\subsubsection{Equilibrium phase-behavior of different molecular pitches}
\begin{figure} [!htb]
	\centering
	\includegraphics[width= 0.9\linewidth]{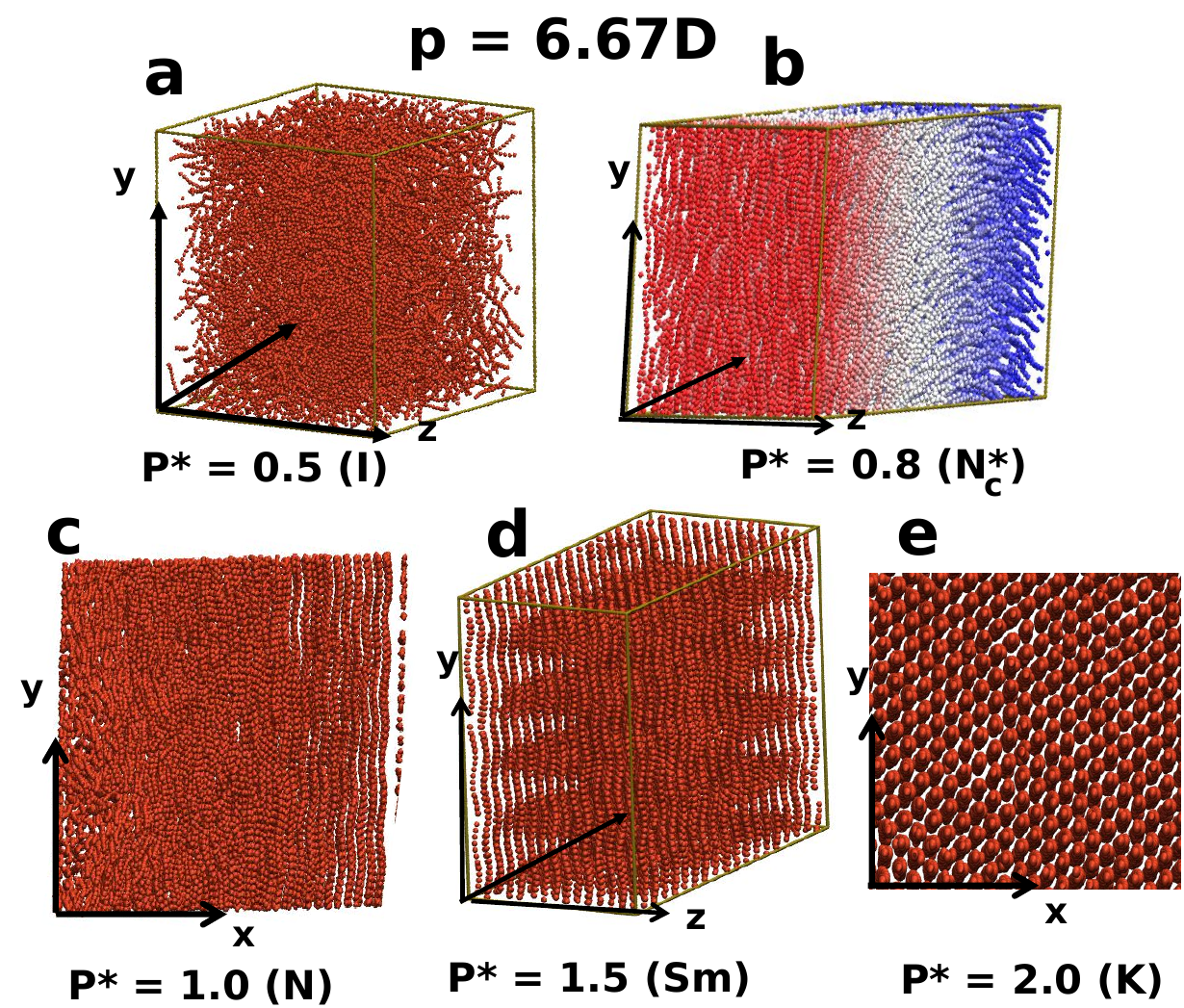}
	\caption{The equilibrium configurations for the system of soft helical particles with molecular pitch $p=6.67D$ (the value of $D$ is taken to be $1.0$), obtained using wall in the simulations, with increasing pressure. a) The isotropic, b)cholesteric, c) nematic, d) smectic and e) crystal phases are obtained at reduced pressure $P^* = 0.5,0.8,1.0,1.5$ and $2.0$ respectively, where $P^* = P/k_{B}T$. In the cholesteric structure (b), particles are colored according to their distance from the helical axis ($\hat{x}$), to show the relative rotation of the layers along the helical axis. } \label{fig:with_wall_ordered_snpshots}
	
\end{figure}

\begin{figure*} [!htb]
	\centering
	\includegraphics[width= 0.8\linewidth]{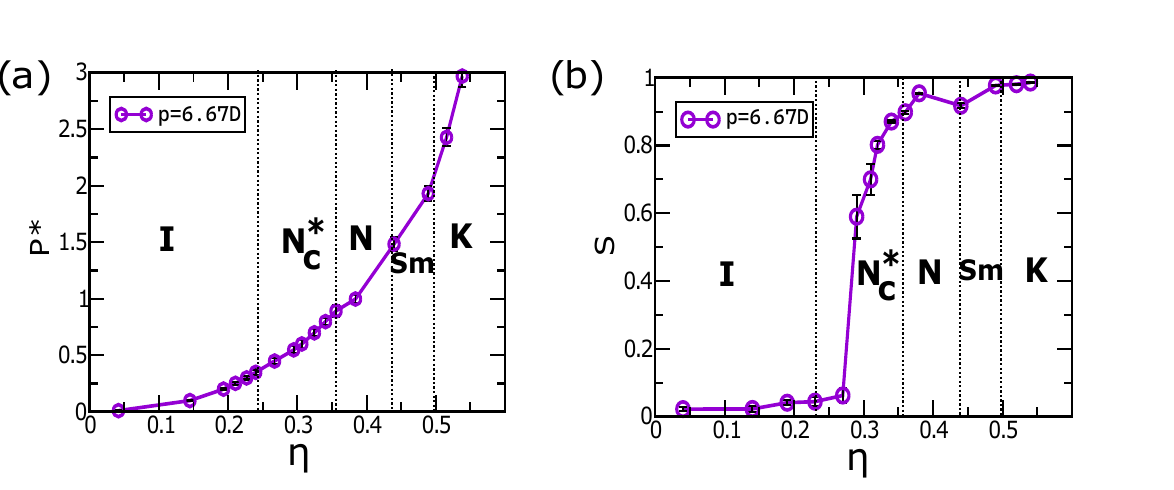}
	\caption{(a) Equation of state (b) Nematic order parameter $S$ vs packing fraction $\eta$ for the pitch $p = 6.67D$. 
 Here, we get five equilibrium phases at the temperature $T^* = 1.0$ for the given range of pressure ($P^*$): isotropic (I), cholesteric ($N_c^*$), nematic (N), Smectic (Sm) and crystal (K). The dotted lines indicate the coexistence region near the phase transition points. 
 }\label{fig:eos-nem-p6.67}
	
\end{figure*}

\begin{figure*} [!htb]
	\centering
	\includegraphics[width= 1\linewidth]{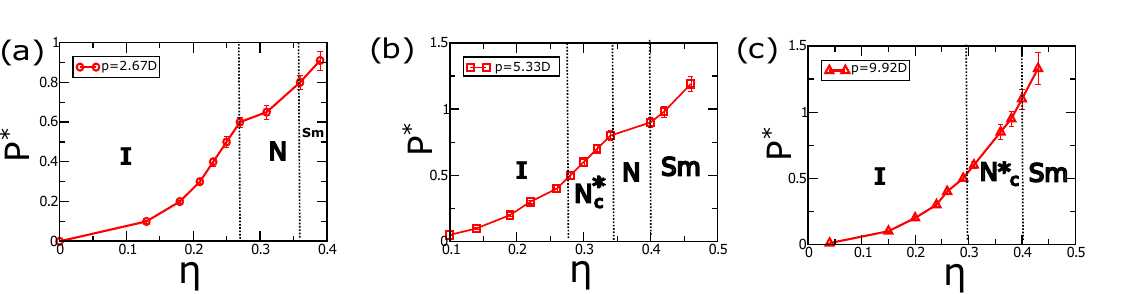}
	\caption{Equation of state for different molecular pitches (a) $p = 2.67D$, (b) $p = 5.33D$, and (c) $p = 9.92D$ with $r = 0.20D$ at the temperature $T^* = 1.0$. We find three stable phases for the given range of pressure for $p = 2.67D$: isotropic, nematic, smectic; four stable phases for pitch $p = 5.33D$: isotropic, cholesteric, nematic and smectic; three stable phases for $p = 9.92D$: isotropic, cholesteric, and smectic. Dotted lines indicate phase boundaries.} \label{fig:eos-all-p}
	
\end{figure*}

We compute the equilibrium phase diagram of left-handed helices of different molecular pitches $p = 2.67D, 5.33D, 6.67D$ and $9.92D$ with $r = 0.2D$ at the temperature $T^{*} = 1$. In Fig \ref{fig:with_wall_ordered_snpshots}, we show the different equilibrium phases found for the system of molecular chirality $p=6.67 D$, which are (i) isotropic (I) (ii) cholesteric ($N_c^*$) (iii) nematic (N) (iv) smectic (Sm) and v) crystal (K), and present the phase diagram in Fig \ref{fig:eos-nem-p6.67}. I-N phase transition is determined by calculating the nematic order parameter $S$ as shown in Fig. \ref{fig:eos-nem-p6.67}(b). The cholesteric phase is quantified by calculating the cholesteric pitch, $P_{chol}$ as discussed in the section \ref{sec:level1}. We calculate the equation of state for other pitches as well, as shown in Fig. \ref{fig:eos-all-p}. For $p = 5.33D$, we observe four stable phases for the given range of densities: (i) isotropic (ii) cholesteric (iii) nematic and (iv) smectic.
For $p = 9.92D$, we observe three stable phases for the given range of densities: (i) isotropic (ii) cholesteric and (iii) smectic. The $I-N_c^*$ transition happens at the packing fraction $\eta^* \approx 0.30-0.32$. Here, we have not found the existence of the nematic phase. From the cholesteric phase, the system undergoes phase transition directly to the smectic phase as can be seen in Fig. \ref{fig:eos-all-p} (c). In the case of molecular pitch $p=2.67D$, we observe three stable phases for the given range of densities: (i) isotropic (ii) nematic and iii) Smectic. Here, we don't find emergence of the cholesteric phase as mentioned in Fig. \ref{fig:eos-all-p}(a). 

\subsubsection{Emergence of cholesteric phase} \label{subsubsec:emergence-of-cholesteric-phase}

\begin{figure}[]
    \centering
    \includegraphics[width=0.9\linewidth]{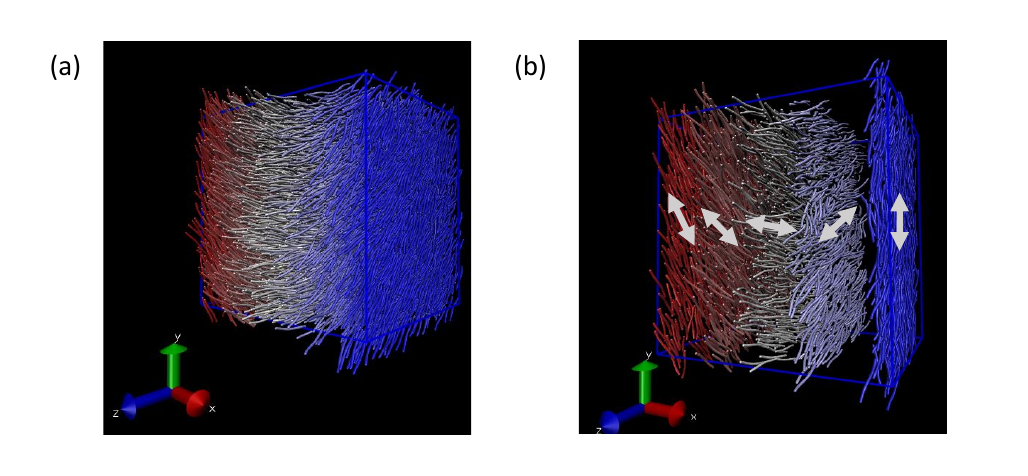}
    \caption{(a) Configuration of the system in the cholesteric phase for the molecular pitch $p = 9.92D$ at the packing fraction $\eta^* = 0.40$. 
    (b) Configuration of the system divided into a number of slabs. The local nematic director corresponding to each slab is shown by the white arrow.
    The particles are colored according to their distance along the x-axis: red, blue and gray colors denote the position of the two edges and center of the box respectively along the $\hat{x}$ direction. 
    We see that, the local nematic director is rotating around the helical axis ($\hat{x}$).  For convenience, only 5 slabs are shown.}
    \label{fig:chol_snapshotpres1}
\end{figure}
\begin{figure}[!htb]
    \centering
    \includegraphics[width= 0.8\linewidth]{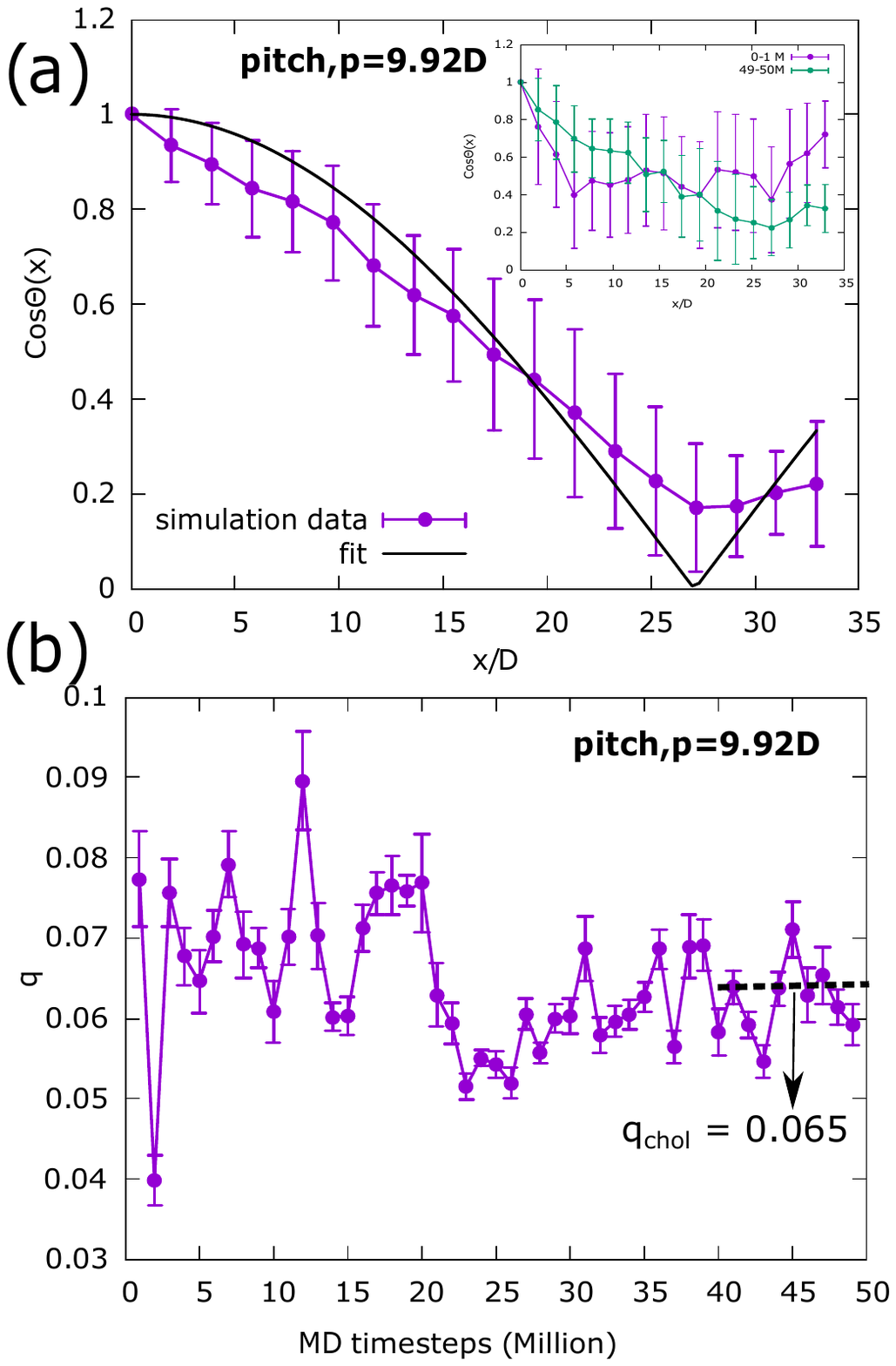}
    \caption{Determination of the cholesteric pitch $P_{chol}$ for the molecular pitch $p=9.92D$ at the packing fraction $\eta=0.40$. 
    (a) The plot of $Cos\theta(x)=|\hat{n(x)}\cdot\hat{n(0)}|$ as a function of the distance along the helical axis $\hat{x}$, where $\hat{n}(x)$ is the local nematic director of each slab located at $x$. 
    The purple line shows the simulation data, and the black smooth curve indicates the linear fitting with the function $|Cos(qx)|$, from which the cholesteric pitch $P_{chol} = 2\pi/q$ is calculated. Inset shows the variation of $Cos\theta(x)$ for different time windows (purple curve for initial $0-1 M$ steps, green curve for system at $49-50 M$ steps). 
    (b) The calculated value of $q$ with time for reduced pressure $P^* = 0.8$. 
    We calculate the cholesteric pitch $P_{chol}$ by taking average of $q$ for the time window 40-50M, where the fluctuation in the values of $q$ is reduced.}
    \label{fig:pitch_with_time-compact-p9.92-pres1}
    \end{figure}

\begin{figure}[]
    \centering
    \includegraphics[width= 0.95\linewidth]{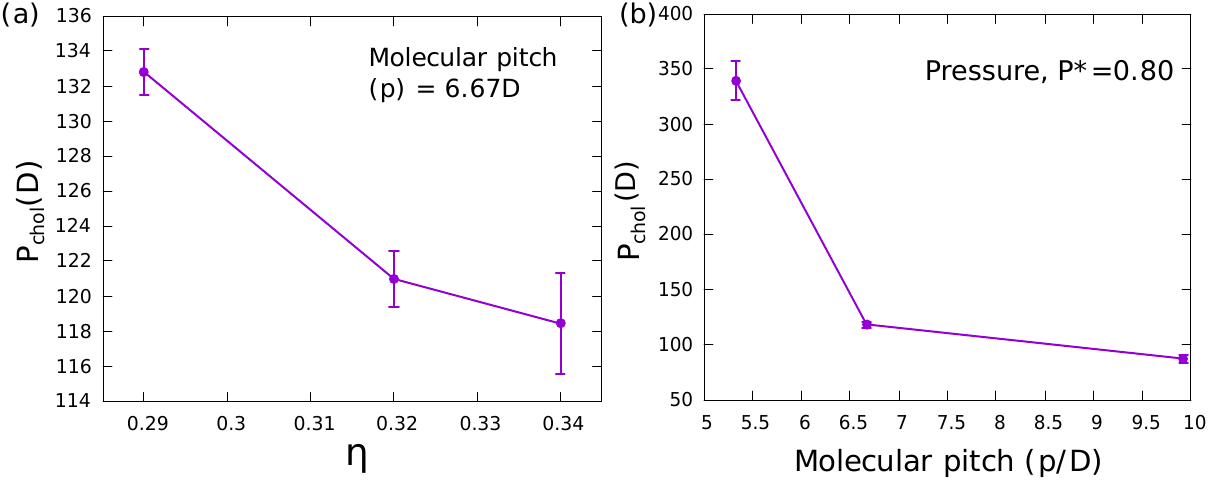}
    \caption{(a) The dependence of the macroscopic pitch of the  cholesteric phase $P_{chol}$ with packing fraction $\eta$ for the molecular pitch $p = 6.67D$. (b) The dependence of the $P_{chol}$ on the molecular pitch $p$ at a specific pressure $P^* = 0.8$. We see that, for a given molecular pitch $p$, $P_{chol}$ decreases with the increase of the pressure and for a given pressure, $P_{chol}$ decreases with the increase of the molecular pitch $p$.}
    \label{fig:pitch-combined}
\end{figure}

We mainly focus on the emergence of the cholesteric phase $N_{c}^{*}$ for the aforementioned molecular pitches. In Fig. \ref{fig:chol_snapshotpres1}, the configuration of a cholesteric phase has been shown for the pitch $p = 9.92D$ for pressure $P^{*} = 1$, temperature $T^* = 1$ and packing fraction $\eta = 0.4$. In Fig. \ref{fig:pitch_with_time-compact-p9.92-pres1}(a), we plot $Cosine$ of the angle $\theta$ (the angle between the local nematic director of each slab $\hat{n}(x)$ with that of the slab at $x = 0$ i.e. $\hat{n}(0)$) with the distance along the helical axis $x$ and fit it with the function $|Cos(qx)|$. 
We obtain the magnitude of the cholesteric pitch to be $P_{chol}= 125.66 D$ with $q = 0.05$.  Our calculated values  are in quantitative agreement with  the cholesteric pitch reported by Kolli \textit{et al.} \cite{hima-JCP-2017} ($q = 0.055$ ) for hard helices using MC simulations. 
Interestingly, we find cholesteric phase for pitches, $p = 6.67D, 5.33D$ which was predicted theoretically earlier by  Dijkstra \textit{et al.}\cite{marjolein-PRE-2014}.
In fig \ref{fig:pitch_with_time-compact-p9.92-pres1}(b) we showed the time variation of the pitch for a system with $p=9.92$ at $P^* = 0.8$.
For a given molecular pitch $p$, $P_{chol}$ decreases with density, as shown in Fig. \ref{fig:pitch-combined}(a). 
The magnitude of $P_{chol}$ mainly depends on the molecular pitch, $p$ of the helix as shown in Fig. \ref{fig:pitch-combined}(b).
We find that, at a certain packing fraction $\eta$, $P_{chol}$ decreases with the increase of the molecular pitch, $p$. For very low molecular pitch ($p = 2.67D$), the cholesteric phase becomes unstable, and the system shows screw-like nematic phase. This observation is consistent with the results   reported by Kolli  \textit{et al.}\cite{hima-softmatter-2014,hima-JCP-2017} for hard helices.

\subsection{Activity induced phase separation}

\begin{figure*} [!htb]
	\centering
	\includegraphics[width= 0.8\linewidth]{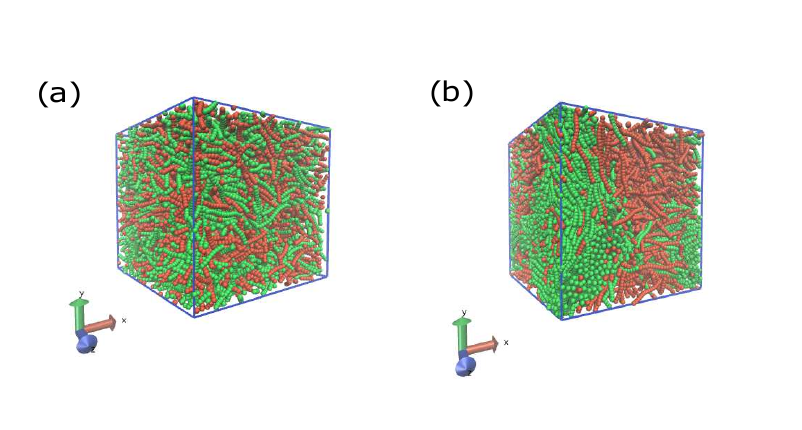}
	\caption{ (a) Equilibrium configuration of $N = 2000$ soft helical particles of pitch $p = 6.67D$ at the state point ($\eta = 0.20$ $T^* = 1$ in the
absence of activity $\chi = 0$. Both hot (red) and cold (green) particles are well mixed at the same temperature. (b) Steady-state configuration after phase separation at $\chi = 9$. It is clearly visible
that cold particles are segregated and ordered, whereas the surrounding hot particles are disordered.} \label{fig:snapshot-active}
	
\end{figure*}

\begin{figure*} [!htb]
	\centering
	\includegraphics[width= 1\linewidth]{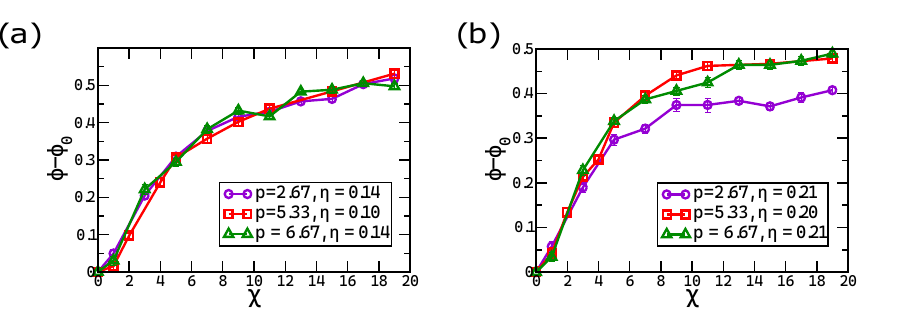}
	\caption{Density order parameter $\phi$ vs activity $\chi$ for different molecular pitches $p$ (in units of $D$) at the packing fractions (a) $\eta = 0.14$ and (b) $\eta = 0.20$. $\phi_{0}$ indicates the magnitude of the density order parameter in the equilibrium system, i.e. at $\chi = 0$. } \label{denop}
	
\end{figure*}

After obtaining the equilibrium phase behavior of soft helical particles, we investigate how these phase behaviors are affected by scalar activity using the two-temperature model. Starting from a homogeneous isotropic structure, we find the emergence of 2-TIPS (2-temperature induced phase separation) phenomena, in which we observe a local phase separation between hot and cold particles that increases with activity till a well-defined interface is formed. In Fig. \ref{fig:snapshot-active}, we show the configuration of a phase separated structure. The extent of phase separation is quantified by calculating density order parameter $\phi$ in the same way as mentioned in our earlier works \cite{tips_mips,Siva,Active-SRS,Ons-JC}. Here, we briefly mention it for completeness. We divide the simulation box into a number of sub-boxes $N_{box}$, and for each sub box we calculate the number difference of hot $n_{h}$ and cold $n_{c}$ particles divided by the total number of particles in each box. We then calculate $\phi$ by averaging over all the sub-boxes and over a sufficient number of configurations to get stable statistics.
\begin{equation}
\phi=\dfrac{1}{N_{box}} \left\langle\sum_{i=1}^{N_{box}}\frac{\lvert n^{i}_{c}-n^{i}_{h}\rvert}{(n^{i}_{c}+n^{i}_{h})} \right\rangle _{ss}, \label{phase-sep-eq}
\end{equation} 
where, $\langle...\rangle_{ss}$ denotes the steady state average over the configurations. $\phi$ is offset by its initial value in the equilibrium at $\chi = 0$ \cite{Active-SRS}. 
In Fig. \ref{denop}, we plot $\phi$ as a function of activities $\chi$ for different molecular pitches at the packing fractions $\eta = 0.14, 0.20$. These packing fractions correspond to the isotropic phase in equilibrium for the given pitches, as mentioned in Fig. \ref{fig:eos-nem-p6.67} and Fig. \ref{fig:eos-all-p}. We observe that phase separation does not depend much on the molecular pitches. For each packing fraction, the magnitude of $\phi$ increases with the activities up to a certain value, then saturates, ensuring macroscopic phase separation between hot and cold particles.

\subsection{Activity induced liquid crystal ordering}

\begin{figure*} [!htb]
	\centering
	\includegraphics [width= 1\linewidth]{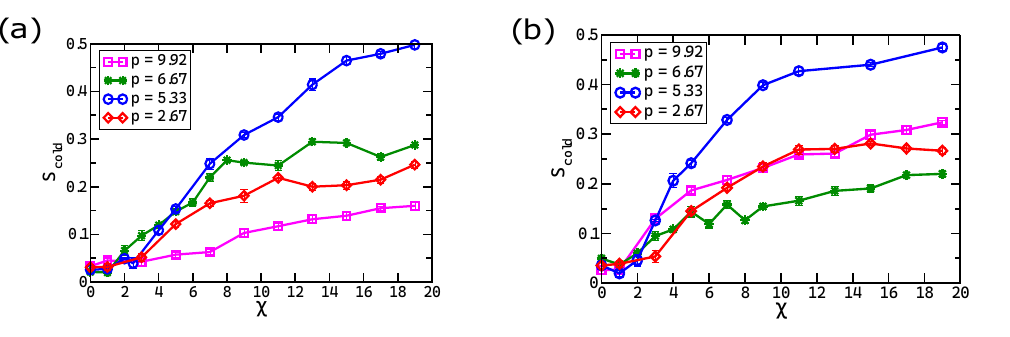}
	\caption{Nematic order parameter of the cold particles $S_{cold}$ as a function of activity $\chi$ for different molecular pitches $p$ at the packing fractions (a) $\eta = 0.14$ and (b) $\eta = 0.20$. $p$ is expressed in units of $D$.} \label{nem-cold}
	\end{figure*}

\begin{figure} [!htb]
	\centering
	\includegraphics[width= 0.7\linewidth]{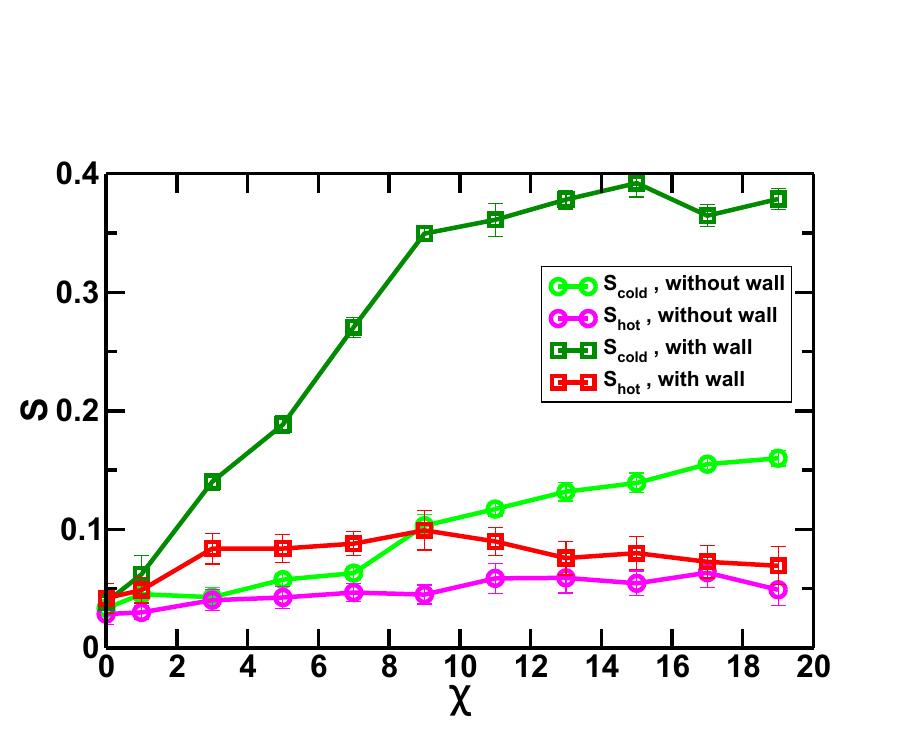}
	\caption{Nematic order parameter of the cold $S_{cold}$ and hot $S_{hot}$ particles
 with and without wall with activity $\chi$ for the pitch $p = 9.92D$ at the packing fractions $\eta = 0.14$. We see that $S_{cold}$ increases in presence of the wall. } \label{nem-wall}
	
\end{figure}

The hot particles receive a sustained energy flux from the hot bath and transfer it to the cold particles through collisions. Cold particles reject the excess energy to the cold bath resulting in a steady flow of energy throughout the system \cite{Ons-JC}. After the phase separation, hot particles exert an extra kinetic pressure on the interface, resulting in an ordering transition in the cold zone \cite{Active-SRS, Ons-JC}. Starting from a homogeneous isotropic state, we observe cold particles undergoing phase transition from isotropic to nematic and then other higher-ordered states, while hot particles remain in the isotropic state with less density. This is interesting as the cold particles show an ordering transition even in the absence of wall, which is not possible in the equilibrium case as discussed in section III-B(1). Presumably, one can think that the interface in the active system provides an alignment direction to the cold particles, thus, playing the role of the wall in the equilibrium. 

We quantify the ordering transition by calculating the nematic order parameter of the cold particles $S_{cold}$, as shown in Fig. \ref{nem-cold}. We find that $S_{cold}$ increases upon increasing activity for the given molecular pitches; however, at a certain activity, $S_{cold}$ have different values for different molecular pitches indicating different local ordering of the cold particles. For example, at a specific value of $\chi$, $p = 5.33D $ exhibits higher ordering for both densities, while, $p = 9.92D $ exhibits lower ordering at $\eta = 0.14$ and $p = 6.67D$ exhibits lower ordering at $\eta = 0.20$. In the case of $p = 9.92D$, we perform the two-temperature simulation in presence of the wall, where we find better ordering. In Fig. \ref{nem-wall}, we observe that for $p = 9.92D$, $S_{cold}$ increases drastically in the presence of the wall than that of the system without a wall at the same activity. Thus, wall promotes higher ordering in active cases also.

\subsection{Breakdown of cholesteric phase}

\begin{figure} [!htb]
	\centering
	\includegraphics[width= 1\linewidth]{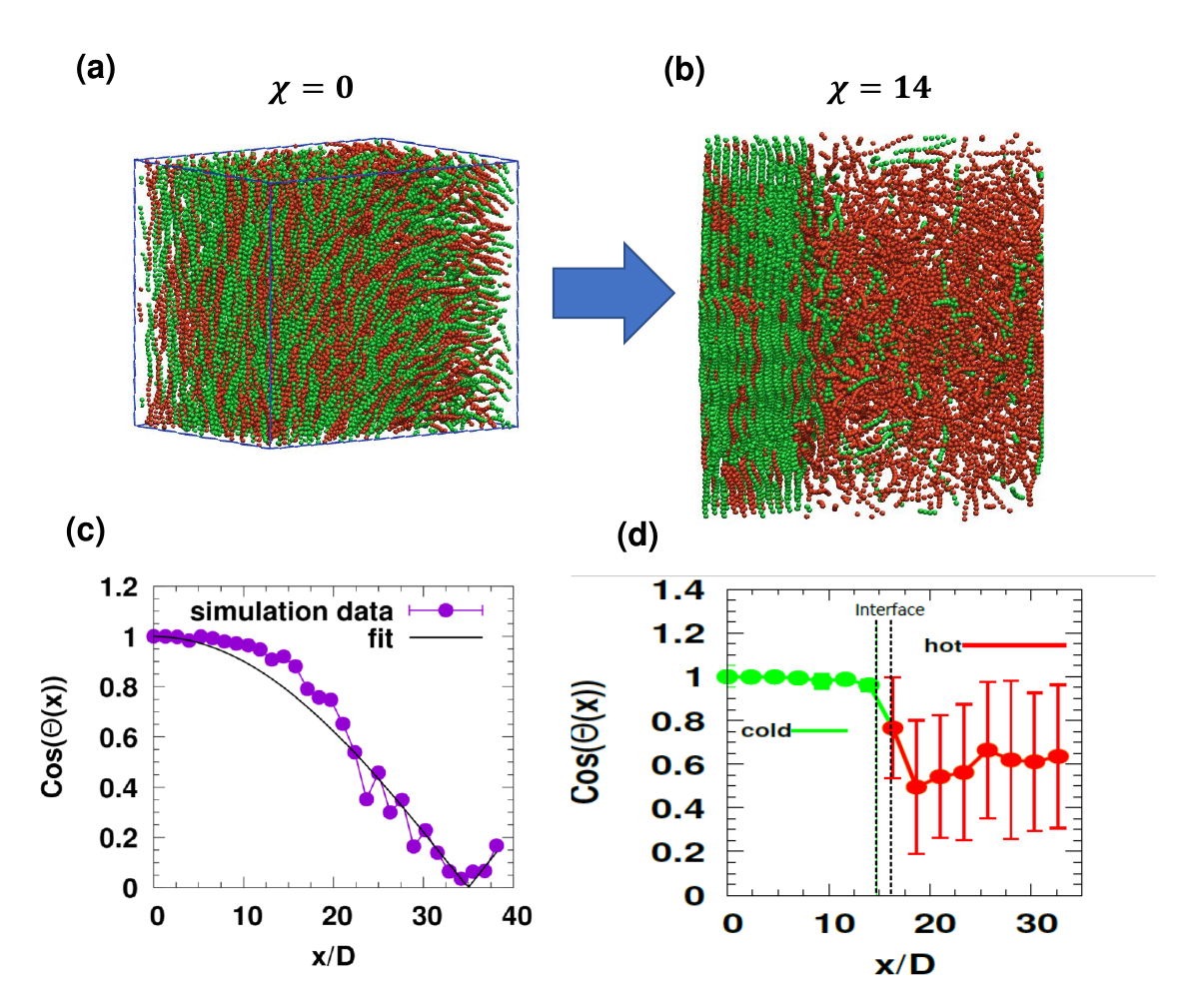}
	\caption{Breakdown of cholesteric phase for the molecular pitch $p = 6.67D$ at pressure $P^* = 0.55$. (a) initial configuration at $\chi = 0$ that shows cholesteric phase. (b) Configuration of the phase-separated system at $\chi = 14$. Calculation of cholesteric pitch $P_{chol}$ for (c) $\chi = 0$ and (d) $\chi = 14$. In equilibrium, $P_{chol}= 133D$. $cos \theta(x)$ is homogeneous in the segregated zone, as shown in figure (d).} \label{p6.67-active}
	
\end{figure}

Starting from a cholesteric phase, we observe that activity destabilizes the cholesteric phase by driving the cold particles through a phase transition to the next higher-ordered state and the hot particles to a state of less order. This result is consistent with our previous work on SRS, as mentioned in Ref. \cite{Active-SRS,Ons-JC}.
In Fig. \ref{p6.67-active}(b), we see that, for $p = 6.67D$, the cold particles form a smectic phase while the hot zone shows an isotropic structure. The breakdown of the $N_c^*$ phase is quantified by plotting $Cos \theta(x)$ with the distance perpendicular to the hot-cold interface ($\theta$ is the relative angle between the local nematic directors at a slab $x$ and at $x=0$, see \ref{subsubsec:emergence-of-cholesteric-phase}). Fig. \ref{p6.67-active}(c) shows the behaviour of $Cos\theta (x)$ for the initial cholesteric structure, whereas in Fig \ref{p6.67-active} (d) we see that $Cos \theta(x)$ is homogeneous in the segregated cold dominated zone indicating the disappearance of the cholesteric phase. At high enough activity, the smectic ordering in the cold zone eventually transforms into a chiral crystal phase.

Similar phenomena have been observed when we start from an initial cholesteric phase for a system of particles with microscopic pitch $p=9.92D$. Starting with a cholesteric phase at pressure $P^* = 0.9$, we introduce activity in the system through the two temperature model. We observe phase separation between active and passive particles. Though there is a fraction of cold particles trapped inside hot zone and fraction of hot particles trapped inside cold zone. The ordering of cold or passive particles gradually increase with activity, but the cholesteric ordering cannot persist in the region, whereas  the hot particles form a disordered isotropic phase (Fig \ref{fig:p9.92-active-Th6-cholbreak}). A smectic layering is shown to occur in the cold domain. The isotropic phase forms at the interior of the system, whereas the smectic domain forms close to the walls. Therefore, we conclude that the disappearance of the cholesteric phase with the introduction of the scalar activity is a universal phenomena as far as different microscopic pitches concerned.


\section{\label{sec:level3}Conclusion and Future Outlook}

In this work, we examined the effect of chirality on the phase behavior of equilibrium and active helical particles driven by two-temperature scalar activity. We first construct the equation of state of soft helical rod-like particles of various intrinsic chiralities using molecular dynamics (MD) simulation. We observe the occurrence of ordered nematic ($N$) and cholesteric ($N_{c}^{*}$) phases in the presence of the wall, for the various values of molecular pitch $p$, namely $p=9.92D,6.67D$ and $5.33D$. Next, we introduce activity in the system through the two-temperature model. Starting from a homogeneous isotropic ($I$) phase, we find that activity causes the hot and cold particles to phase-separate and cold particles to undergo an ordering transition even in the absence of a wall. This observation unveiled that the hot-cold interface in the active system acts as a wall in the equilibrium by giving the cold particles some alignment directions. However, starting from a cholesteric phase, activity destabilizes the $N_{c}^{*}$ phase by inducing smectic ordering in the cold zone while isotropic structure in the hot zone. The smectic ordering in the cold zone eventually transforms to a chiral crystal phase at high enough activity.\\
The future outlook of this work is as follows: in equilibrium, another interesting prospect of the work would be to determine the handedness of the macroscopic cholesteric phase and how they depend on the density and chirality of the constituent particles. A full phase diagram using molecular dynamics simulation would be interesting to explore. The existence of a cholesteric phase for systems with microscopic pitches which were earlier not reported also calls for extensive study on the equilibrium properties of such systems. 

\bibliography{ref}

\appendix*
\renewcommand{\thefigure}{A\arabic{figure}}
\setcounter{figure}{0}
\section{Absence of ordered phases without wall }

\begin{figure} [!htb]
	\centering
	\includegraphics[height= 0.8\linewidth]{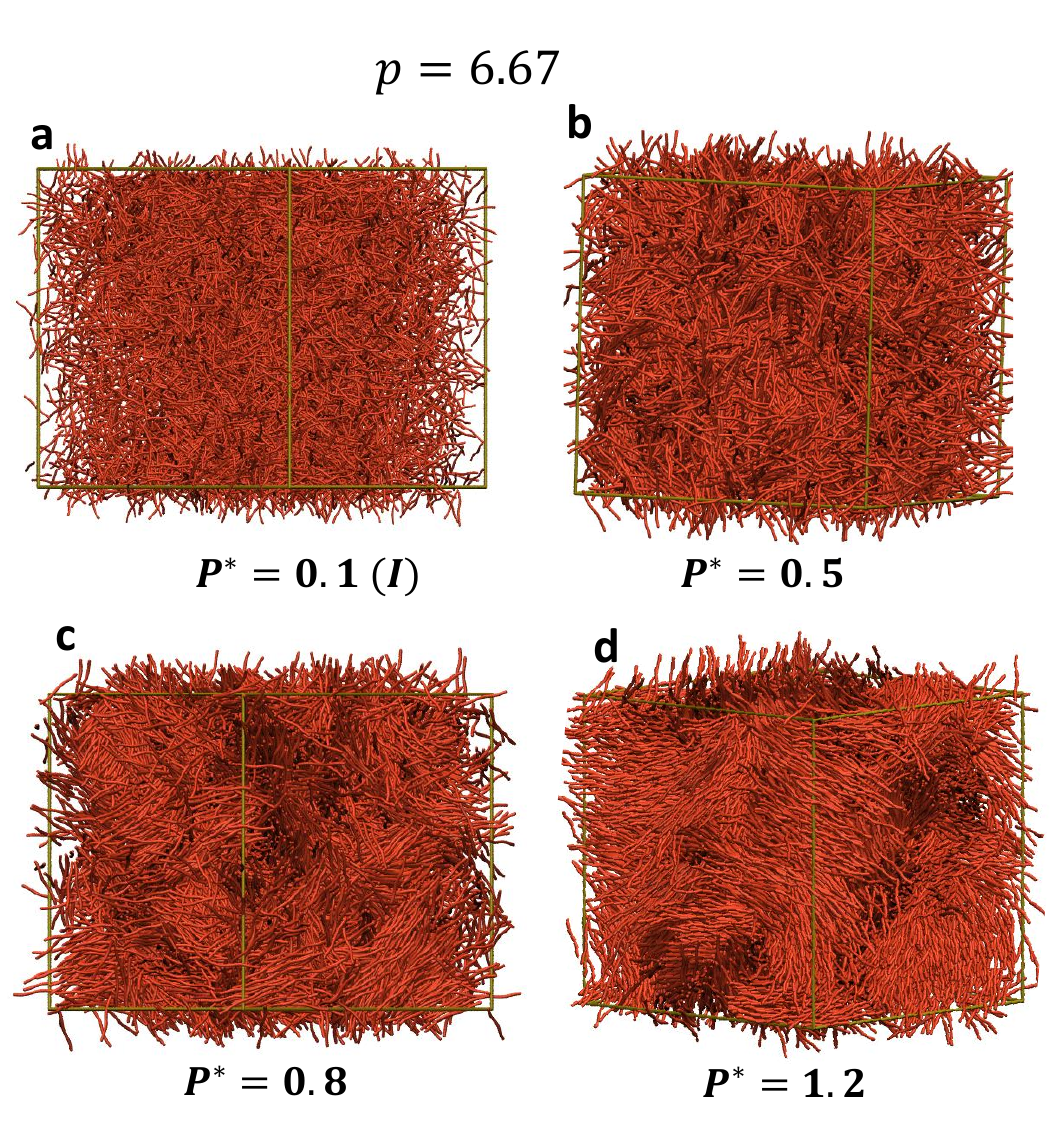}
	\caption{Representation of the configurations obtained for a system of chiral particles with $p=6.67D$ (the value of $D$ is taken to be $1.0$), showing a) Isotropic phase $(I)$ at low pressure $P^* = 0.1$, and b) c) d) jammed state at pressure $P^* = 0.5,0.8$ and $1.2$ respectively. The pressure is expressed in reduced units $P^* = \frac{P}{k_BT}$} \label{fig:without_wall_jammed_snapshots}
	
\end{figure}
 

Simulating the system of soft helical rods with periodic boundary conditions exhibits jammed state at high densities, although the low density isotropic phases is observed in the system. In order to avoid the jammed state, we have used walls in our simulations, that help in the formation of the cholesteric phase within the simulation timescale, by offering an homeotropic alignment to the particles. Fig \ref{fig:without_wall_jammed_snapshots} shows the snapshots of such jammed states at different pressures.

\section{Breakdown of Cholesteric phase}


\begin{figure} [!htb]
	\centering
	\includegraphics[height= 1\linewidth]{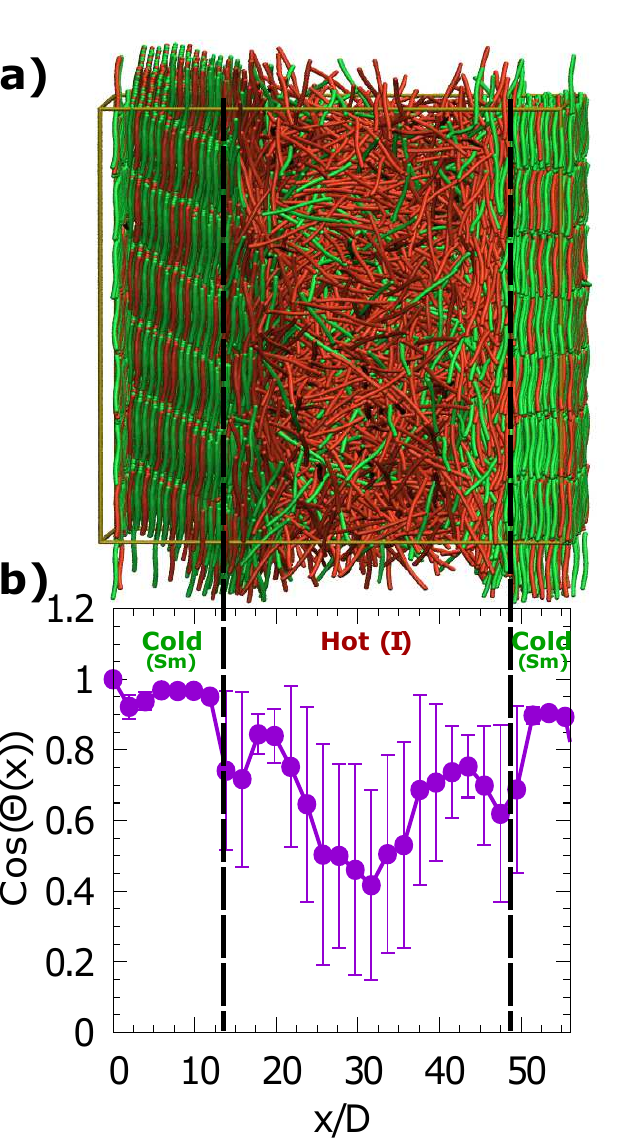}
	\caption{ For molecular pitch $p = 9.92D$: a) The final configuration of the system at $\chi = 5$, the cold particles form ordered smectic phase near the walls, whereas the hot particles form the isotropic phase at the interior. b) The variation of $Cos(\theta)$ with $x$ for the same system. The regions near the wall show lesser fluctuation in the value of $Cos(\theta)$ due to smectic ordering, whereas isotropic region shows larger fluctuation. Also the overall variation of $Cos(\theta)$ doesnot show the behaviour typical to cholesteric phase, indicating the breakdown of the initial cholesteric phase. } \label{fig:p9.92-active-Th6-cholbreak}
	
\end{figure}
 

 Starting from an equilibrated cholesteric phase for a system of soft helix of microscopic pitch $p = 9.92$ at reduced pressure $P^* = 0.9$, we use the two temperature model to observe the fate of the cholesteric phase. Phase separation between the hot and cold particles takes place. Our results show that when activity is introduced, the hot particles form an isotropic phase while the cold particles exhibit an ordered phase that is shown to be smectic in the observed range of activities. Therefore the initial cholesteric phase is breaking down under the influence of temperature difference.


\end{document}